\documentclass{article}
\usepackage{spconf}

\usepackage{amsmath,amssymb,bbm}
\usepackage{cite}
\usepackage[breaklinks=true,letterpaper=true,bookmarks=false,hidelinks]{hyperref}
\usepackage{cleveref}
\usepackage{url}
\usepackage{mathptmx}
\usepackage{amsmath}
\usepackage{bm}
\usepackage{graphicx}
\usepackage[caption=false]{subfig}
\usepackage[utf8]{inputenc} 
\usepackage[T1]{fontenc}    
\usepackage{hyperref}       
\usepackage{url}            
\usepackage{booktabs}       
\usepackage{amsfonts}       
\usepackage{nicefrac}       
\usepackage{microtype}      
\usepackage{xcolor}         

\usepackage{caption}

\usepackage{algorithm}
\usepackage{algorithmic}
\usepackage{color,soul}
\usepackage{booktabs}
\usepackage{array}
\newcolumntype{L}[1]{>{\raggedright\let\newline\\\arraybackslash\hspace{0pt}}m{#1}}
\newcolumntype{C}[1]{>{\centering\let\newline\\\arraybackslash\hspace{0pt}}m{#1}}
\newcolumntype{R}[1]{>{\raggedleft\let\newline\\\arraybackslash\hspace{0pt}}m{#1}}
\mathchardef\mhyphen="2D

\title{Synt++: Utilizing Imperfect Synthetic Data to Improve Speech Recognition}
 
\name{
\begin{tabular}{c} Ting-Yao Hu$^{\star}$ \qquad Mohammadreza Armandpour$^{\dagger}$ \qquad Ashish Shrivastava$^{\star}$ \\
   Jen-Hao Rick Chang$^{\star}$ \qquad Hema Koppula$^{\star}$ \qquad Oncel Tuzel$^{\star}$\end{tabular}
\thanks{$^\dagger$Work done during summer internship at Apple. Emails: $^{\star}$\{tingyao\_hu, ashish.s, jenhao\_chang, hkoppula, otuzel\}@apple.com, $^\dagger$armand@stat.tamu.edu}
\vspace{-0.15in}
}
\address{
$^\star$Apple \;\;\;\;\;\;\;\; $^\dagger$Texas A\&M University
}

\graphicspath{{./Figures/}}

\begin{document}

\maketitle
 
\begin{abstract}

With recent advances in speech synthesis, synthetic data is becoming a viable alternative to real data for training speech recognition models. 
However, machine learning with synthetic data is not trivial due to the gap between the synthetic and the real data distributions.  Synthetic datasets may contain artifacts that do not exist in real data such as structured noise, content errors, or unrealistic speaking styles. 
Moreover, the synthesis process may introduce a bias due to uneven sampling of the data manifold. 
We propose two novel techniques during training to mitigate the problems due to the distribution gap: (i) a rejection sampling algorithm and (ii) using separate batch normalization statistics for the real and the synthetic samples. We show that these methods significantly improve the training of speech recognition models using synthetic data. We evaluate the proposed approach on keyword detection and Automatic Speech Recognition (ASR) tasks, and observe up to $18\%$ and $13\%$ relative error reduction, respectively, compared to naively using the synthetic data.



%
\end{abstract}

\begin{keywords}
Data augmentation, speech recognition, synthetic data
\end{keywords}
\vspace{-0.1in}

\section{Introduction}
\label{sec: intro}

\begin{figure}[t]
    \centering
    \includegraphics[width=.75\linewidth]{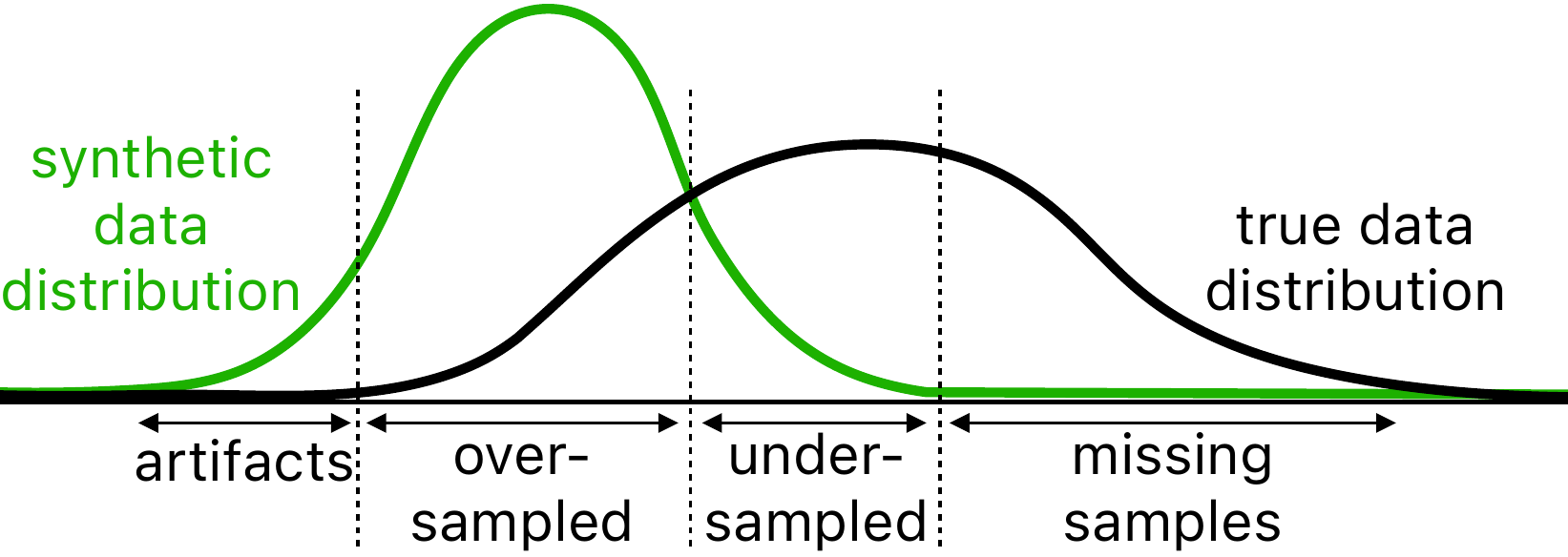} \\
    \vspace{-0.1in}
    \caption{The joint distributions of speech and corresponding text. The gap between synthetic and true data distributions can be partitioned into four regions. See text for details.}
    
    \label{fig:motivation}
\vspace{-0.2in}
\end{figure}

Collecting and annotating datasets for training speech recognition models is hard and expensive.
An alternative approach is to generate a synthetic dataset using speech synthesis models~\cite{vasquez2019_melnet,Li2019NeuralSS,Ma2019iclr,jia2018transfer,mist_hu_2020,style_tokens_wang18h, style_eq_chang_2021}.
However, training recognition models with synthetic data is not trivial. 
%
%
Figure~\ref{fig:motivation} shows an illustration of the joint distribution of speech signals and corresponding text labels -- black is the true data distribution and green is the synthetic data distribution.
In general, we do not have access to the true data distribution, but have finite sample approximation through a collected real dataset.
Although the quality of synthetic speech (synthesized via a controllable generative model) has improved significantly, there is still a gap between the synthetic and the true data distributions. 
We can characterize this gap into 4 different regions as shown in Figure~\ref{fig:motivation}: (i) artifacts (e.g. structured noise, speech/content mismatch, unrealistic styles in synthetic data) where the synthesized samples are outside the support of true  distribution, (ii) over-sampled region where the synthetic distribution has more mass and (iii) under-sampled region where the synthetic distribution has less mass compared to the true distribution, and (iv) missing samples where the synthetic distribution has zero mass inside the support of the true distribution (e.g. missing speaking styles, accents). 

In this paper, we present Synt++, an improved algorithm to utilize synthetic data for training speech recognition models.
Synt++ combines two novel techniques.
The first technique is a rejection sampling algorithm~\cite{prml_book,azadi2018discriminator,pmlr_grover_2018} that modifies the sampling process of the synthesis model (the generative model) to make the synthesized data distribution closer to the true data distribution.
Specifically, we train a discriminator function to classify samples as real or synthetic. 
We then use the probability of real measure predicted by this discriminator to stochastically accept/reject synthesized samples.
The rejection sampling addresses the first three problems discussed above by rejecting artifacts, and under/over sampling data points in over/under-sampled regions, respectively. 
Note that the rejection sampling cannot correct for the missing samples since the synthesis model does not have support and, hence, cannot synthesize the samples in this region. 

The second technique accounts for the remaining gap during training by using separate Batch Normalization (BN) statistics for real and synthetic data. BN is a commonly used technique to normalize the features during training to address covariate shift and improve convergence.
However, the BN statistics estimated using a combination of synthetic and real data becomes biased due to the distribution gap (e.g., in Figure~\ref{fig:motivation}, the mean of the combined real/synthetic distribution is shifted left and the variance is larger than the true distribution). 
To prevent this problem, we estimate real and synthetic data BN statistics separately, and utilize only the real BN statistics during inference. 
Moreover, this process helps to bridge the domain gap since synthetic and real data are normalized separately to have similar statistics. 
\begin{figure}[t]
    \centering
    \includegraphics[width=.75\linewidth]{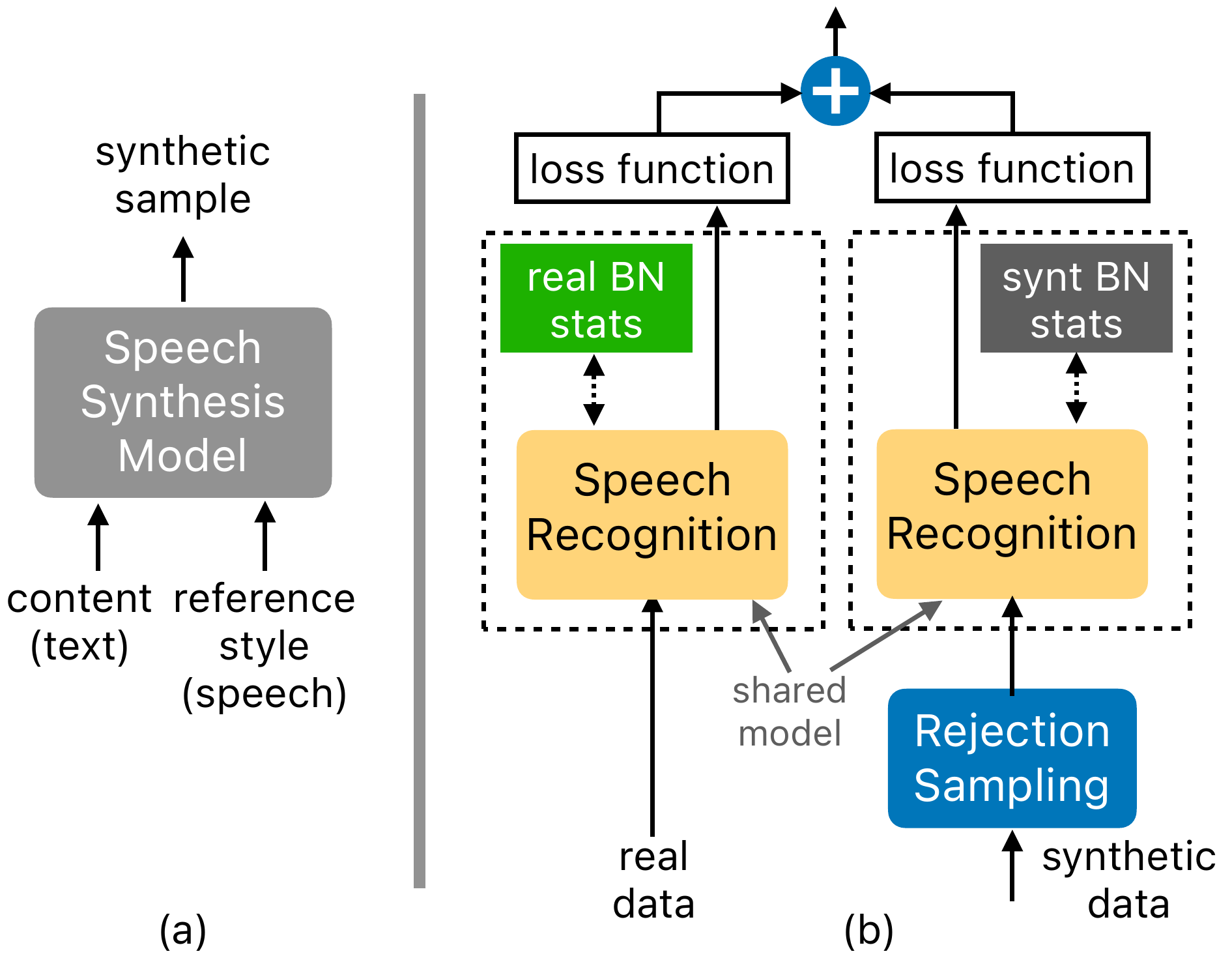} \\
    \vspace{-0.1in}
    \caption{Method overview. (a) To synthesize data, we use a controllable generative model~\cite{style_eq_chang_2021} that takes a text and a reference speech as input and utters the text in the style of the given speech. (b) During training of the recognition model, we use rejection sampling and separate BN statistics to address the distribution gap between the real and the synthetic data.}
    \vspace{-0.3cm}
    \label{fig:overview}
\end{figure}

\vspace{-0.2cm}
\section{Related Work}
State-of-the-art neural Text-to-Speech (TTS) architectures, such as Tacotron 2 \cite{shen2018natural} and FastSpeech 2 \cite{ren2020fastspeech}, are capable of generating speech indistinguishable from human speech. Controllable speech synthesis extends the TTS models beyond modeling the voice of a single or a few speakers to the entire style space of speech  \cite{style_tokens_wang18h,mist_hu_2020,style_eq_chang_2021,deepvoice}. Using controllable synthesis models, we can expand the acoustic variations of synthetic speech or use a bigger text corpus to add data for unseen content.
%


Previous works have used TTS models to improve the training of speech recognition models. 
One group of work relies on the TTS models and self/semi-supervised learning techniques to exploit unpaired speech and text data in the ASR training process \cite{tjandra2019end,baskar2019semi,hori2019cycle,hayashi2018back,baskar2021eat}.
Another group leverages TTS models to create more training data when the amount of real data is limited \cite{rossenbach2020generating,du2020speaker,rosenberg2019speech,mimura2018leveraging,chen2020improving}.
Some other works have focused on specific tasks, such as out-of-vocabulary word recognition \cite{zheng2021using}, and keyword detection \cite{werchniak2021exploring,lin2020training}.
These methods, however, do not address the imperfections of the speech generated by the synthesis models. 

The idea of utilizing synthetic data to improve a recognition model has also been explored in the computer vision domain.
In \cite{shrivastava2017learning}, the authors proposed to refine synthetic images via adversarial training to bridge the distribution gap. 
In \cite{antoniou2017data}, the authors showed that a generative model could be trained to augment images of unseen classes.
In \cite{Xie_2020_CVPR}, authors use adversarial examples to improve image recognition.

\vspace{-0.3cm}
\section{Method}
\vspace{-0.1cm}
\subsection{Controllable Speech Generative Model}
\vspace{-0.1cm}
Controllable speech generative models can generate speech in a given style, where the content is specified by an input text. 
As shown in Figure~\ref{fig:motivation}, for effective learning with synthetic data, we need a generative model that has \emph{as small gap as possible} with the true distribution. 
Therefore, the generative model should be able to synthesize any text in a wide range of speaking styles in various environmental conditions such as background noise, microphone response, etc. without introducing artifacts to the signal. 
To this end, we utilize a recently introduced auto regressive controllable generative model called Style Equalization~\cite{style_eq_chang_2021} that enables learning a controllable generative model in the wild (thousands of speakers recorded in various conditions) without requiring style labels. 
As shown in Figure~\ref{fig:overview}(a), the model takes a text and a reference style speech as input and utters the text in the style of the given reference. 
In this model, the style is broadly defined, not only the speaker’s voice characteristics but also all aspects of the
reference speech sample, including background noise, echo, microphone response, etc., which are necessary to have a smaller distribution gap (particularly missing samples). 
This model also allows sampling styles from a prior distribution that is important for sampling new styles not contained in the dataset. 
We refer readers to~\cite{style_eq_chang_2021} for a detailed description of this model.

\vspace{-0.3cm}
\subsection{Rejection Sampling}


\label{method:rej_sampling}
Let $\boldsymbol x$ be a speech signal and $y$ be the corresponding token label, and $p_d(\boldsymbol x, y)$, $p_g(\boldsymbol x, y)$ denote the joint true and synthetic data distributions, respectively.
We use the synthesizer as proposal distribution and use rejection sampling to make it closer to the true distribution.
Following standard rejection sampling~\cite{prml_book}, a synthetic data $(\boldsymbol x, y)$ is accepted with probability $\frac{p_d(\boldsymbol x, y)}{M p_g(\boldsymbol x, y)}$, where scalar $M > 0$ is selected such that $M p_g(\boldsymbol x, y) >p_d(\boldsymbol x, y), \forall (\boldsymbol x, y)$ in domain of $p_d$, i.e., $M = \max_{\boldsymbol x, y} \frac{p_d(\boldsymbol x, y)}{p_g(\boldsymbol x, y)}$.
Azadi et. al.~\cite{azadi2018discriminator}  proposed a rejection sampling method to match the distribution of the images from a generative adversarial network to the true distribution; however, their algorithm does not consider the joint distribution of $(\boldsymbol x, y)$ that is needed to train downstream recognizers.

To approximate the ratio $\frac{p_d(\boldsymbol x, y)}{p_g(\boldsymbol x, y)}$, we train a discriminator, $D(\boldsymbol x, y)$,  that takes a speech signal and the corresponding text and produces probability of $(\boldsymbol x, y)$ coming from true distribution, i.e., $D(\boldsymbol x, y) \in [0, 1]$. 
When $D$ is trained to its optimal value $D^*$, it satisfies  (\cite{gan_goodfellow_2014,azadi2018discriminator,pmlr_grover_2018})
\vspace{-0.3cm}
\begin{align}
    D^*(\boldsymbol x, y) = \frac{p_d(\boldsymbol x, y)}{p_d(\boldsymbol x, y) + p_g(\boldsymbol x, y)},
    \label{eq:D_star_equals_pd_div_pd_plus_pg}
\end{align}
which can be re-arranged to yield
\begin{align}
r(\boldsymbol x, y) := \frac{p_d(\boldsymbol x, y)}{p_g(\boldsymbol x, y)} =  \frac{D^*(\boldsymbol x, y)}{1-D^*(\boldsymbol x, y)}.
\label{eq:pd_div_pg}
\end{align}
To train $D$, we minimize
\vspace{-0.3cm}
\begin{align}
\mathcal L_D = \mathbb E_{(\boldsymbol x, y) \sim p_d} (\log (D(\boldsymbol x, y))) + \mathbb E_{(\boldsymbol x, y) \sim p_g} (1 - \log (D(\boldsymbol{x}, y))), \nonumber
\end{align}
which is equivalent to training a two class classification model, where the positive class (label $1$) corresponds to the real samples and the negative class (label $0$) corresponds to the synthetic samples. 
Note that, we model the joint speech and label distribution, therefore $D$ should also utilize whether the speech signal matches the token labels. 
To this end, we use a pre-trained ASR model, $A$ to predict the token labels, $\hat{y}$, of the speech signal, i.e., $\hat{y} = A(\boldsymbol x)$. 
Then, we characterize the discrepancy between $\hat{y}$ and $y$ by a 5-dimensional feature vector $\Phi(\boldsymbol x,y)$, whose elements include the cross-entropy (CE) loss, connectionist temporal classification loss, word error rate (WER), and number of tokens in $y$ and $\hat{y}$.
Finally, we train a two-layer DNN, $D_{\boldsymbol \theta}$, that takes $\Phi(\boldsymbol x,y)$ as input and produces the probability of the sample being real 
\begin{align}
    D(\boldsymbol x, y):=  D_{\boldsymbol \theta}(\Phi(\boldsymbol x, y)),
    \label{eq:D_score}
\end{align}
where $\boldsymbol \theta$ denotes the parameters of the DNN.


Practically, we need to compute $M$ and address the fact that the rejection sampling can be slow to select a fixed number of samples.
First, we estimate an initial value of $M$ by computing the maximum value of $p_d / p_g$  (using $D^*$ and Eq.~\ref{eq:pd_div_pg}) over a few (approximately $200$) generated samples. 
Following~\cite{azadi2018discriminator}, to compute the acceptance probability for each generated sample, $(\boldsymbol x, y)$, we first compute the discriminator score $D^*(\boldsymbol x, y)$ and then set $M \leftarrow \max(M, {p_d(\boldsymbol x, y) / p_g(\boldsymbol x, y)}) = \max(M, \frac{D^*(\boldsymbol x, y)}{1-D^*(\boldsymbol x, y)})$).
We stop the sampling process after accepting $N$ synthetic samples. Note that, the proposed rejection sampling technique applies only to the domain where the generative model has support and ignores the missing samples region shown in Figure~\ref{fig:motivation}.


\subsection{Double Batch Normalization Statistics}


\label{method:dbl_bn}
A BN layer~\cite{batchnorm_ioffe15} computes the mean and the standard deviation of the intermediate features and normalizes these features by subtracting the mean and dividing by the standard deviation within a batch. 
During training, the model uses means and variances computed over the samples in a mini-batch.
A running average of the BN statistics, which we call running statistics, is computed to be used during inference.

Naively using BN will update the running statistics with samples from both the real and the synthetic distributions and, due to the distribution gap, the estimates will be biased (i.e. the estimated BN statistics will not match the true data distribution).
Instead, we form the mini-batches consisting of only the real samples or only the synthetic samples, and compute two sets of running statistics -- one from the real mini-batches and the other from the synthetic mini-batches. 
Since the test data (during inference) consists of only real samples, we discard the synthetic statistics and use only the real statistics. Note that, we share the affine parameters of the BN between synthetic and real batches. 

The proposed technique is similar to Xie et al.~\cite{Xie_2020_CVPR} that used auxiliary BN layers in an adversarial learning framework where images augmented with adversarial perturbations were used to improve model accuracy. In contrast, we use separate BN statistics to improve model training with synthetic data.

%
\vspace{-0.3cm}
\section{Experiments}
\label{sec:experiment}
 \subsection{Automatic Speech Recognition}

For ASR experiments, we use the LibriSpeech dataset~\cite{panayotov2015librispeech}, which contains $1,000$ hours of speech from public domain audiobooks.
Following the standard protocol, we evaluate our method on the full training set (LibriSpeech 960h) and a subset containing clean speech with only US English accents (LibriSpeech 100h).
To generate synthetic datasets, we use Style Equalization~\cite{style_eq_chang_2021} that is trained with LibriTTS dataset \cite{zen2019libritts}.
LibriTTS is a subset of LibriSpeech dataset and contains $555$ hours of speech data.
To study the effect of model size on synthetic data training, we train two ASR models -- a large model with 116 million parameters (Large 116M) and a smaller model with 22 million parameters (Regular 22M)

\textbf{ASR model: }
We use the implementation from ESPNet \cite{watanabe2018espnet}) for an end-to-end ASR model.
The ASR model is composed of a conformer-based encoder \cite{gulati2020conformer} and a transformer-based decoder \cite{Li2019NeuralSS}. 
Setting the encoder dimension to $144$ and $512$, we construct the regular (22M) and the large (116M) models.
We apply SpecAugment \cite{park2019specaugment} to all speech samples to further enhance the acoustic diversity.
The model checkpoint of each epoch is saved, and the final model is produced by averaging the 10 checkpoints with the best validation accuracy. 
All ASR models are evaluated without a language model.

\textbf{Baselines: }  For synthetic-only training, we compare naively using synthetic data with the proposed rejection sampling (Section~\ref{method:rej_sampling}).
Note that when the training data consists of only synthetic samples, we do not have any real data to compute the running BN statistics.
Hence, we do not use the double BN when training only with synthetic data.
For joint (real,synt) training, our baseline is naively adding the synthetic data to training.
We compare this baseline with the proposed method (real, synt++) where we use the rejection sampling algorithm described in Section~\ref{method:rej_sampling} and also use separate BN statistics for the real and the synthetic samples (Section~\ref{method:dbl_bn}).

\textbf{Synthetic data with the same text corpus as the real data:}
First, we study the effectiveness of the synthetic data when the synthetic samples are used to increase the acoustic variation in the training data. 
To this end, we use the same training corpus as the real data in LibriSpeech 960h.
Specifically, for each sentence $y_i$ in the training dataset, we take a random real training sample, $\boldsymbol x_j$, as the style reference, and generate synthetic sample $\boldsymbol x'_i$. For both Synt and Synt++ (using rejection sampling), we obtain a synthetic dataset containing 960 hours of  speech.


The WERs of the recognition models are reported in Table~\ref{tb:librispeech_960} on the two official test sets (test-clean and test-other).
Synt++ gives significant improvement over naively adding synthetic data to the training (Synt).

\textbf{Synthetic data with extended corpus: }
We study the effect of synthesizing speech from an extended text corpus in Table~\ref{tb:librispeech_100}. 
In addition to LibriSpeech 960h, we also report the results when we have a smaller real dataset (LibriSpeech 100h). 
For LibriSpeech 100h, we use the text corpus of LibriSpeech 960h and synthesize 960 hours of synthetic data.
For LibriSpeech 960h dataset, we double its text corpus by adding the text from the language model corpus of LibriSpeech and synthesize 960 hours of synthetic data. In both cases, extended corpus significantly improves the performance. Particularly, for the smaller dataset, we observe $48\%$ relative  reduction in WER ($7.7\%$ to $4.0\%$), compared to real-only training. 


\textbf{Ablation study: }
To isolate the effect of the rejection sampling and the double BN, we conduct an ablation study in Table~\ref{tb:librispeech_960_ablation} on LibriSpeech 960h dataset.
As seen in this table, both the rejection sampling and the double BN statistics are important to effectively utilize the synthetic samples.


\begin{table}[t]
\footnotesize
\centering
\setlength{\tabcolsep}{3pt}
\begin{tabular}{@{}C{1.3cm}C{1.2cm}C{1.3cm}C{1.3cm}C{1.15cm}C{1.3cm}@{}}
  \toprule
   & Real & Synt & Synt++ & Real, \;Synt & Real, Synt++\\ 
  \midrule
   \multicolumn{6}{l}{\hspace{-0.18cm} Regular (22M)}\\\hline
   ~~test-clean &  $3.7\pm0.14$ & $7.3\pm0.17$ & $7.0\pm0.05$ & $3.4\pm0.05$ & $\mathbf{3.2}\pm0.08$  \\
   ~~test-other &  $9.5\pm0.12$ &$21.0\pm0.12$ & $20.0\pm0.17$  & $9.3\pm0.21$ & $\mathbf{8.5}\pm0.05$  \\ \midrule
 \multicolumn{6}{l}{\hspace{-0.18cm} Large (116M)}\\\hline
   ~~test-clean & $2.9\pm0.05$ & $6.3\pm0.05$ & $5.4\pm0.12$ & $3.0\pm0.05$ & $\mathbf{2.6}\pm0.05$  \\
   ~~test-other & $6.9\pm0.08$ & $18.2\pm0.17$ & $17.2\pm0.17$  & $7.4\pm0.17$ & $\mathbf{6.5}\pm0.05$  \\
 \bottomrule
\end{tabular}
\vspace{-0.3cm}
\caption{ASR results on LibriSpeech 960h dataset. The reported numbers are percentage WER (lower is better). 
Synt++ significantly improves training with synthetic data.}
\vspace{-0.1cm}
\label{tb:librispeech_960}
\end{table}



\begin{table}[t]
\centering
\scriptsize
\setlength{\tabcolsep}{0.1pt}
\begin{tabular}{@{}C{1.3cm}C{1.2cm}C{1.2cm}C{1.2cm}C{0.1cm}C{1.2cm}C{1.2cm}C{1.2cm}@{}}
  \toprule
  & \multicolumn{3}{c}{\footnotesize{LibriSpeech 100h}} & \phantom{abc}& \multicolumn{3}{c}{\footnotesize{LibriSpeech 960h}} \\
   \cmidrule{2-4}\cmidrule{6-8}
   & \footnotesize{Real} & \footnotesize{Real, \;Synt\;} & \footnotesize{Real, Synt++}  & \phantom{abc} & \footnotesize{Real} & \footnotesize{Real, \;Synt\;} & \footnotesize{Real, Synt++}\\ 
  \midrule

    test-clean & \scriptsize{$7.7\pm 0.08$} & \scriptsize{$4.3\pm 0.08$} & \scriptsize{$\mathbf{4.0} \pm0.08$}  & \phantom{abc} & \scriptsize{$2.9 \pm0.05$} & \scriptsize{$2.7 \pm 0.08$} & \scriptsize{$\mathbf{2.4} \pm0.05$}  \\
     test-other & \scriptsize{$20.9\pm 0.64$} & \scriptsize{$13.2\pm 0.21$} & \scriptsize{$\mathbf{13.2} \pm0.08$} & \phantom{abc} &  \scriptsize{$6.9 \pm0.08$} &  \scriptsize{$7.0 \pm0.25$} &  \scriptsize{$\mathbf{6.3} \pm0.05$} \\
 \bottomrule
\end{tabular}
\vspace{-0.2cm}
\caption{Effect of using an extended text corpus.}
\label{tb:librispeech_100}
\vspace{-0.3cm}
\end{table}

\begin{table}[t]
\footnotesize
\centering
\setlength{\tabcolsep}{2pt}
\begin{tabular}{@{}L{1.7cm}C{1.4cm}C{1.4cm}C{0.4cm}C{1.4cm}C{1.4cm}@{}}
  \toprule
& \multicolumn{2}{c}{Regular (22M)} & \phantom{abc}& \multicolumn{2}{c}{Large (116M)} \\
 \cmidrule{2-3}\cmidrule{5-6}
   & test-clean & test-other & \phantom{abc} & test-clean & test-other \\ \midrule
   Real, Synt &$3.4\pm0.05$ & $9.3\pm0.21$ & \phantom{abc} & $3.0\pm0.05$ & $7.4\pm0.17$  \\ 
   + rejection  &$3.4\pm0.05$ & $9.0\pm0.12$ & \phantom{abc} & $2.8\pm0.02$ & $7.0\pm0.05$  \\ 
   + dbl BN & $3.3\pm0.02$ & $8.5\pm0.02$ & \phantom{abc} & $2.7\pm0.05$ & $6.7\pm0.02$  \\
   + rej. + dbl BN &$\mathbf{3.2}\pm0.08$  & $\mathbf{8.5}\pm0.05$  & \phantom{abc} & $\mathbf{2.6}\pm0.05$ & $\mathbf{6.5}\pm0.05$  \\ 
 \bottomrule
\end{tabular}
\vspace{-0.2cm}
\caption{Ablation study on the ASR task with LibriSpeech 960h dataset.}
\vspace{-0.3cm}
\label{tb:librispeech_960_ablation}
\end{table}

\subsection{Keyword Detection}

We also evaluate the proposed approach on keyword detection task using the Speech Command dataset~\cite{speechcommandsv2}.
This dataset consists of $35$ keywords from multiple speakers, and samples from various background noises.
For each keyword, we split the train and test subset based on the speaker identities, i.e. training and test samples did not have common speakers.
We chose $3$ keywords (`down', `no', `stop') to study the improvements with synthetic data.
Each of these keywords contains approximately $2500$ training samples and $800$ test samples.
We set up the keyword detection task as a 2-class classification problem, where the positive class is one of the keywords and the negative class contains all the other $34$ keywords and the background noise.
Each sample consists of approximately 1 second of audio, which is converted to a mel-spectrogram before giving it as input to a  ResNet model with approximately 6k parameters.
The model is trained with cross-entropy loss.

\textbf{Sampling synthetic data: } For each keyword, we generated $10$k samples by randomly selecting the reference styles from the real dataset.
For the positive class, we generated an additional $100$k samples.
For both synthetic and real training, we added approximately $5$k samples from background noise to the negative class.
Since the detection model predicts the keyword probability (unlike a token sequence in the ASR task), we use only the CE loss in Equation~\ref{eq:D_score} to compute $D^*$. 

\textbf{Metric:} Since keyword detection is a 2-class problem, we can compute false reject rate (FRR) and false accept rate (FAR) at different thresholds. 
The detection error tradeoff (DET) curve is commonly used for keyword detection tasks~\cite{e2e_dnnhmm_icassp_2021}.
Since lower FRR is preferred for keyword detection, we report average FAR for FRR values over the range of $0\mhyphen 5\%$, which is equivalent to the area under the curve (AUC) of the DET curve over the FRR range of $0\mhyphen 5\%$.

\textbf{Results:} We report the average FAR for keywords in Table~\ref{tb:speechcmd} and observe significant improvements of the proposed approach, compared to naively using the synthetic data.

\begin{table}[t]
\footnotesize
\centering
\setlength{\tabcolsep}{3pt}
\begin{tabular}{@{}C{1.2cm}C{1.2cm}C{1.2cm}C{1.3cm}C{1.1cm}C{1.3cm}@{}}
  \toprule
  keywords & Real & Synt & Synt++ & Real, Synt & Real, Synt++\\ 
  \midrule
   `down' & $9.6\pm0.8$ & $11.2\pm0.1$ & $8.9\pm0.9$ & $5.6\pm0.2$ & $\mathbf{5.0}\pm0.2$  \\
   `no' & $13.5\pm0.8$ & $14.5\pm0.7$ &$11.0\pm0.2$ & $7.8\pm0.4$ & $\mathbf{6.4}\pm0.4$  \\
   `stop' & $3.5\pm0.3$ & $11.2\pm0.6$ & $9.6\pm0.4$ &$2.1\pm0.3$ & $\mathbf{1.5}\pm0.1$  \\
 \bottomrule
\end{tabular}
\vspace{-0.3cm}
\caption{Keyword detection results on Speech Command dataset. The reported numbers are average false accept rate in percentage (lower is better). }
\vspace{-0.5cm}
\label{tb:speechcmd}
\end{table}

\vspace{-0.3cm}
\section{Conclusions}
We presented Synt++, a new algorithm to train speech recognition models using synthetic data. We applied Synt++ to ASR and keyword detection tasks and obtained significant improvements over training with real-only and real+synthetic datasets.


\vspace{0.2cm}
\setlength{\parindent}{0pt}
\textbf{Acknowledgement:}
We are grateful to our colleagues Russ Webb and Barry Theobald for their valuable inputs.

\footnotesize
\bibliography{mybib}

\begin{thebibliography}{10}

\bibitem{vasquez2019_melnet}
Sean Vasquez and Mike Lewis,
\newblock ``Melnet: A generative model for audio in the frequency domain,''
\newblock {\em arXiv preprint arXiv:1906.01083}, 2019.

\bibitem{Li2019NeuralSS}
Naihan Li, Shujie Liu, Yanqing Liu, Sheng Zhao, and Peng Shi,
\newblock ``Neural speech synthesis with transformer network,''
\newblock in {\em AAAI}, 2019.

\bibitem{Ma2019iclr}
Shuang Ma, Daniel Mcduff, and Yale Song,
\newblock ``A generative adversarial network for style modeling in a
  text-to-speech system,''
\newblock in {\em ICLR}, 2019.

\bibitem{jia2018transfer}
Ye~Jia, Yu~Zhang, Ron Weiss, Quan Wang, Jonathan Shen, Fei Ren, Patrick Nguyen,
  Ruoming Pang, Ignacio~Lopez Moreno, Yonghui Wu, et~al.,
\newblock ``Transfer learning from speaker verification to multispeaker
  text-to-speech synthesis,''
\newblock in {\em Proc. NIPS}, 2018.

\bibitem{mist_hu_2020}
Ting-Yao Hu, Ashish Shrivastava, Oncel Tuzel, and Chandra Dhir,
\newblock ``Unsupervised style and content separation by minimizing mutual
  information for speech synthesis,''
\newblock in {\em Proc. ICASSP}, 2020.

\bibitem{style_tokens_wang18h}
Yuxuan Wang, Daisy Stanton, Yu~Zhang, RJ-Skerry Ryan, Eric Battenberg, Joel
  Shor, Ying Xiao, Ye~Jia, Fei Ren, and Rif~A. Saurous,
\newblock ``Style tokens: Unsupervised style modeling, control and transfer in
  end-to-end speech synthesis,''
\newblock in {\em Proc. ICML}, 2018.

\bibitem{style_eq_chang_2021}
Jen-Hao~Rick Chang, Ashish Shrivastava, Hema Koppula, Xiaoshuai Zhang, and
  Oncel Tuzel,
\newblock ``Style equalization: Unsupervised learning of controllable
  generative sequence models,''
\newblock {\em arXiv preprint arXiv:2110.02891}, 2021.

\bibitem{prml_book}
Christopher~M. Bishop,
\newblock {\em Pattern recognition and machine learning, 5th Edition},
\newblock Information science and statistics. Springer, 2007.

\bibitem{azadi2018discriminator}
Samaneh Azadi, Catherine Olsson, Trevor Darrell, Ian Goodfellow, and Augustus
  Odena,
\newblock ``Discriminator rejection sampling,''
\newblock in {\em Proc. ICLR}, 2019.

\bibitem{pmlr_grover_2018}
Aditya Grover, Ramki Gummadi, Miguel Lazaro-Gredilla, Dale Schuurmans, and
  Stefano Ermon,
\newblock ``Variational rejection sampling,''
\newblock in {\em Proc. AISTATS}, 2018.

\bibitem{shen2018natural}
Jonathan Shen, Ruoming Pang, Ron~J Weiss, Mike Schuster, Navdeep Jaitly,
  Zongheng Yang, Zhifeng Chen, Yu~Zhang, Yuxuan Wang, Rj~Skerrv-Ryan, et~al.,
\newblock ``Natural tts synthesis by conditioning wavenet on mel spectrogram
  predictions,''
\newblock in {\em Proc. ICASSP}, 2018.

\bibitem{ren2020fastspeech}
Yi~Ren, Chenxu Hu, Xu~Tan, Tao Qin, Sheng Zhao, Zhou Zhao, and Tie-Yan Liu,
\newblock ``Fastspeech 2: Fast and high-quality end-to-end text to speech,''
\newblock in {\em Proc. ICLR}, 2021.

\bibitem{deepvoice}
Wei Ping, Kainan Peng, Andrew Gibiansky, Sercan~O Arik, Ajay Kannan, Sharan
  Narang, Jonathan Raiman, and John Miller,
\newblock ``Deep voice 3: Scaling text-to-speech with convolutional sequence
  learning,''
\newblock in {\em Proc. ICLR}, 2018.

\bibitem{tjandra2019end}
Andros Tjandra, Sakriani Sakti, and Satoshi Nakamura,
\newblock ``End-to-end feedback loss in speech chain framework via
  straight-through estimator,''
\newblock in {\em Proc. ICASSP}, 2019.

\bibitem{baskar2019semi}
Murali~Karthick Baskar, Shinji Watanabe, Ramon Astudillo, Takaaki Hori,
  Luk{\'a}{\v{s}} Burget, and Jan {\v{C}}ernock{\`y},
\newblock ``Semi-supervised sequence-to-sequence asr using unpaired speech and
  text,''
\newblock {\em arXiv preprint arXiv:1905.01152}, 2019.

\bibitem{hori2019cycle}
Takaaki Hori, Ramon Astudillo, Tomoki Hayashi, Yu~Zhang, Shinji Watanabe, and
  Jonathan Le~Roux,
\newblock ``Cycle-consistency training for end-to-end speech recognition,''
\newblock in {\em Proc. ICASSP}, 2019.

\bibitem{hayashi2018back}
Tomoki Hayashi, Shinji Watanabe, Yu~Zhang, Tomoki Toda, Takaaki Hori, Ramon
  Astudillo, and Kazuya Takeda,
\newblock ``Back-translation-style data augmentation for end-to-end asr,''
\newblock in {\em IEEE Spoken Language Technology Workshop (SLT)}, 2018.

\bibitem{baskar2021eat}
Murali~Karthick Baskar, Luk{\'a}{\v{s}} Burget, Shinji Watanabe,
  Ramon~Fernandez Astudillo, et~al.,
\newblock ``Eat: Enhanced asr-tts for self-supervised speech recognition,''
\newblock in {\em Proc. ICASSP}, 2021.

\bibitem{rossenbach2020generating}
Nick Rossenbach, Albert Zeyer, Ralf Schl{\"u}ter, and Hermann Ney,
\newblock ``Generating synthetic audio data for attention-based speech
  recognition systems,''
\newblock in {\em Proc. ICASSP}, 2020.

\bibitem{du2020speaker}
Chenpeng Du and Kai Yu,
\newblock ``Speaker augmentation for low resource speech recognition,''
\newblock in {\em Proc. ICASSP}, 2020.

\bibitem{rosenberg2019speech}
Andrew Rosenberg, Yu~Zhang, Bhuvana Ramabhadran, Ye~Jia, Pedro Moreno, Yonghui
  Wu, and Zelin Wu,
\newblock ``Speech recognition with augmented synthesized speech,''
\newblock in {\em ASRU}, 2019.

\bibitem{mimura2018leveraging}
Masato Mimura, Sei Ueno, Hirofumi Inaguma, Shinsuke Sakai, and Tatsuya
  Kawahara,
\newblock ``Leveraging sequence-to-sequence speech synthesis for enhancing
  acoustic-to-word speech recognition,''
\newblock in {\em IEEE Spoken Language Technology Workshop (SLT)}, 2018.

\bibitem{chen2020improving}
Zhehuai Chen, Andrew Rosenberg, Yu~Zhang, Gary Wang, Bhuvana Ramabhadran, and
  Pedro~J Moreno,
\newblock ``Improving speech recognition using gan-based speech synthesis and
  contrastive unspoken text selection.,''
\newblock in {\em Proc. Interspeech}, 2020.

\bibitem{zheng2021using}
Xianrui Zheng, Yulan Liu, Deniz Gunceler, and Daniel Willett,
\newblock ``Using synthetic audio to improve the recognition of
  out-of-vocabulary words in end-to-end asr systems,''
\newblock in {\em Proc. ICASSP}, 2021.

\bibitem{werchniak2021exploring}
Andrew Werchniak, Roberto~Barra Chicote, Yuriy Mishchenko, Jasha Droppo, Jeff
  Condal, Peng Liu, and Anish Shah,
\newblock ``Exploring the application of synthetic audio in training keyword
  spotters,''
\newblock in {\em Proc. ICASSP}, 2021.

\bibitem{lin2020training}
James Lin, Kevin Kilgour, Dominik Roblek, and Matthew Sharifi,
\newblock ``Training keyword spotters with limited and synthesized speech
  data,''
\newblock in {\em Proc. ICASSP}, 2020.

\bibitem{shrivastava2017learning}
Ashish Shrivastava, Tomas Pfister, Oncel Tuzel, Joshua Susskind, Wenda Wang,
  and Russell Webb,
\newblock ``Learning from simulated and unsupervised images through adversarial
  training,''
\newblock in {\em Proc. CVPR}, 2017.

\bibitem{antoniou2017data}
Antreas Antoniou, Amos Storkey, and Harrison Edwards,
\newblock ``Data augmentation generative adversarial networks,''
\newblock {\em arXiv preprint arXiv:1711.04340}, 2017.

\bibitem{Xie_2020_CVPR}
Cihang Xie, Mingxing Tan, Boqing Gong, Jiang Wang, Alan~L. Yuille, and Quoc~V.
  Le,
\newblock ``Adversarial examples improve image recognition,''
\newblock in {\em Proc. CVPR}, 2020.

\bibitem{gan_goodfellow_2014}
Ian Goodfellow, Jean Pouget-Abadie, Mehdi Mirza, Bing Xu, David Warde-Farley,
  Sherjil Ozair, Aaron Courville, and Yoshua Bengio,
\newblock ``Generative adversarial nets,''
\newblock in {\em Proc. NIPS}, 2014.

\bibitem{batchnorm_ioffe15}
Sergey Ioffe and Christian Szegedy,
\newblock ``Batch normalization: Accelerating deep network training by reducing
  internal covariate shift,''
\newblock in {\em Proc. ICML}, 2015.

\bibitem{panayotov2015librispeech}
Vassil Panayotov, Guoguo Chen, Daniel Povey, and Sanjeev Khudanpur,
\newblock ``Librispeech: an asr corpus based on public domain audio books,''
\newblock in {\em Proc. ICASSP}, 2015.

\bibitem{zen2019libritts}
Heiga Zen, Viet Dang, Rob Clark, Yu~Zhang, Ron~J Weiss, Ye~Jia, Zhifeng Chen,
  and Yonghui Wu,
\newblock ``Libritts: A corpus derived from librispeech for text-to-speech,''
\newblock {\em arXiv preprint arXiv:1904.02882}, 2019.

\bibitem{watanabe2018espnet}
Shinji Watanabe, Takaaki Hori, Shigeki Karita, Tomoki Hayashi, Jiro Nishitoba,
  Yuya Unno, Nelson {Enrique Yalta Soplin}, Jahn Heymann, Matthew Wiesner,
  Nanxin Chen, Adithya Renduchintala, and Tsubasa Ochiai,
\newblock ``{ESPnet}: End-to-end speech processing toolkit,''
\newblock in {\em Proc. Interspeech}, 2018.

\bibitem{gulati2020conformer}
Anmol Gulati, James Qin, Chung-Cheng Chiu, Niki Parmar, Yu~Zhang, Jiahui Yu,
  Wei Han, Shibo Wang, Zhengdong Zhang, Yonghui Wu, et~al.,
\newblock ``Conformer: Convolution-augmented transformer for speech
  recognition,''
\newblock {\em arXiv preprint arXiv:2005.08100}, 2020.

\bibitem{park2019specaugment}
Daniel~S Park, William Chan, Yu~Zhang, Chung-Cheng Chiu, Barret Zoph, Ekin~D
  Cubuk, and Quoc~V Le,
\newblock ``Specaugment: A simple data augmentation method for automatic speech
  recognition,''
\newblock {\em Proc. Interspeech}, 2019.

\bibitem{speechcommandsv2}
P.~{Warden},
\newblock ``{Speech Commands: A Dataset for Limited-Vocabulary Speech
  Recognition},''
\newblock {\em ArXiv e-prints}, 2018.

\bibitem{e2e_dnnhmm_icassp_2021}
Ashish Shrivastava, Arnav Kundu, Chandra Dhir, Devang Naik, and Oncel Tuzel,
\newblock ``Optimize what matters: Training dnn-hmm keyword spotting model
  using end metric,''
\newblock in {\em Proc. ICASSP}, 2021.

\end{thebibliography}
\bibliographystyle{IEEEbib}

%
\end{document}